\documentclass[twoside]{ilcws08}
\usepackage[latin1]{inputenc}
\usepackage[dvips]{graphicx,epsfig,color}
\usepackage{wrapfig,rotating}
\usepackage{amssymb,amsmath,array}

\begin{document}
\title{
Dark Matter in the USSM} 
\author{J. Kalinowski,$^1$ S.F. King$^2$ and J.P. Roberts$^3$
\vspace{.3cm}\\
1- Physics Department, University of Warsaw, 00-681 Warsaw, Poland\\
and Theory Division, CERN, CH-1211 Geneva 23, Switzerland
\vspace{.1cm}\\
2- School of Physics and Astronomy, University of Southampton,\\
Southampton, SO17 1BJ, U.K.
\vspace{.1cm}\\
3- Center for Cosmology and Particle Physics,
New York University,\\
New York, NY 10003, USA}

\maketitle

\begin{abstract}
We discuss the  neutralino
dark matter within classes of extended supersymmetric models, referred
to as the USSM, containing one additional SM singlet Higgs
plus an extra $Z'$, together with their superpartners the singlino and
bino'.
\end{abstract}

\section{Introduction}

Models with additional $U(1)$ gauge group provide an elegant solution to the $\mu$  problem of the minimal supersymmetric standard model (MSSM). The
Higgs/Higgsino mass term $\mu \hat H_1\hat H_2$ of the MSSM is replaced by $ \lambda \hat S\hat H_1\hat H_2 $ (with a dimensionless coupling $\lambda$), where the additional
superfield $\hat S$ is  a singlet under the Standard Model (SM).  The vacuum expectation value  $\langle S\rangle$ then not only
dynamically generates a SUSY Higgs/Higgsino mass near the weak scale
as required but also results in an increased Higgs boson mass upper bound which in turn gives
a welcome reduction in electroweak fine tuning. Moreover, by gauging the Abelian $U(1)_X$ symmetry arising from  the $ \lambda \hat S\hat H_1\hat H_2 $ term the
troublesome axion is avoided via the Higgs mechanism resulting in a massive $Z'$ gauge boson \cite{Fayet:1977yc}. The essential additional elements of such a scenario
then consist of two extra superfields relative to those of the MSSM,
namely the singlet superfield $\hat S$ and the $U(1)_X$ gauge
superfield $B'$.  The USSM is then defined as a model with the MSSM superfields
plus these two additional superfields  at the TeV scale. Since it does not include other
superfields necessary for cancelation of  the fermionic $U(1)_X$ gauge anomalies, it has to be considered as a truncation of a complete model. For example, identifying the Abelian gauge group as a subgroup of $E_6$ with complete 27 dimensional
representations of matter down to the TeV scale solves the anomaly problem, while requiring
further the right-handed neutrinos to be singlets  under $U(1)_X$  (for a see-saw
mechanism) defines the theory uniquely as the
E$_6$SSM~\cite{King:2005jy}.

For the USSM  we adopt the charge assignment under the extra $U(1)_X$ as in the E$_6$SSM
and study the physics and cosmology in this simplified setting to learn about crucial
features which will be relevant to any complete model involving an
additional $U(1)_X$ gauge group and a singlet.
This talk, which updates the preliminary results presented previously~\cite{our},
is based on the recent comprehensive analysis of neutralino dark matter in
the USSM in~\cite{Kalinowski:2008iq}, where full details (and an extensive list of references)
can be found.

\section{The USSM model}
Compared to the MSSM the
particle spectrum is extended by
a new CP-even Higgs boson $S$, a gauge boson $Z'$ and two neutral
--inos: a singlino $\tilde{S}$ and a bino' $\tilde{B}'$. Other
sectors are not enlarged but the sfermion scalar potential receives additional D-terms.

With two additional degrees of freedom the neutralino sector cannot be solved analytically any more. However,  since the
original MSSM and the new degrees of freedom are coupled weakly, an approximate analytical solution can be found
following a two-step diagonalization procedure~\cite{Choi:2006fz}. First - the $4\times 4$ MSSM
and the new $2\times 2$ singlino-bino'
sectors are separately diagonalised; second - a block--diagonalization provides approximate solutions. \begin{wrapfigure}{r}{0.5\columnwidth}
\centerline{\includegraphics[width=0.4\columnwidth, height=0.3\columnwidth]{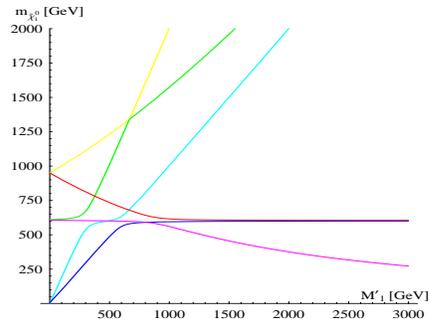}}
\caption{Neutralino mass spectrum.}\label{Fig:masses}
\end{wrapfigure}Since $U(1)_X$ forbids the singlino mass term (in contrast to the NMSSM) the mini see-saw structure of the singlino-bino' mass matrix implies a singlino--dominated lightest neutralino in the limit of large bino' mass.
Figure~\ref{Fig:masses} shows the neutralino mass spectrum as a function of the soft bino' mass parameter $M'_1$ in a GUT-scale  unifying scenario $M'_1=M_1=M_2/2$, where $M_1,\,M_2$ are the $U(1),\,SU(2)$ mass parameters respectively; $\tan\beta=5$, other parameters are chosen so that the effective $\mu=600$ GeV,  the heavy $M_{Z_2}=950$ GeV, and the pseudoscalar Higgs $m_A=500$ GeV.

\section{Results}
The presence of new singlino and bino'
states greatly modifies the phenomenology of the neutralino sector
both at colliders and in cosmology-related processes.
It is informative to consider the general form of the interactions that
arise from the singlino and bino' components of the lightest
neutralino  before presenting our numerical results.

The interactions  of the bino'
component, which is always subdominant due to the see-saw structure of
the neutralino mass matrix, closely mirror those of the bino component, except for the
different coupling constant and charges under the new $U(1)_X$.  On the other hand, the $\lambda \hat{S}\,
\hat{H}_u\hat{H}_d$ term in the superpotential  gives rise to a new type
of neutralino coupling: the lightest neutralino with significant singlino and higgsino components  will couple
strongly to Higgs bosons with a significant $H_u$ or $H_d$ component,
usually the lighter Higgs bosons, $H_{1,2}\text{ and }A$ in the
spectrum.  Moreover, the absence of the singlet cubic term $\tilde
S^3$, in contrast to the NMSSM, implies that the singlino-dominated
LSP needs an admixture of MSSM higgsinos to annihilate through s-channel
Higgs bosons.
Since the singlino component does not interact with the
$SU(2)$ or $U(1)_Y$ gauginos or with fermions, a significant singlino
component in the lightest neutralino will suppress couplings to
$W$ or $Z_1$ bosons and to fermions.

\subsection{Relic density}
The differences
between the MSSM relic density calculation and the USSM calculation
arise through the extension of the particle spectrum and through the
new interactions that are introduced. We have implemented all new
interactions into the {\tt micrOMEGAs}\cite{micromegas} code which
takes full account of all annihilation and coannihilation
processes and calculates their effect whenever they are
relevant.

\begin{wrapfigure}{r}{0.5\columnwidth}
\centerline{\includegraphics[width=0.45\columnwidth]{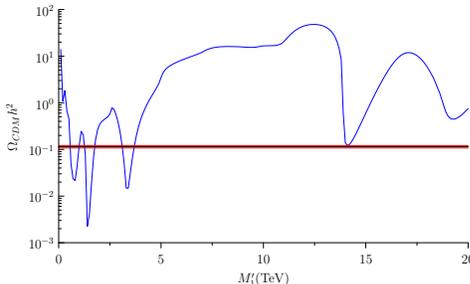}}
\caption{The relic density across varying $M_1'$}\label{Fig:relic}
\end{wrapfigure}
Figure~\ref{Fig:relic} shows the relic density as a function of $M'_1$.
Ignoring initially the resonance effects,  a general trend in
the relic density from a large value at low $M_1'$, down to a lower
value at around $M_1'=0.75$~TeV and then back to larger values at high
$M_1'$ is easily understood as this follows the evolution of the
LSP from bino through higgsino  to
singlino, see Figure~\ref{Fig:masses}.

Beyond this general behavior there are  interesting resonance
structures as  the LSP mass first increases with the $M'_1$ increase
reaching a maximum of $\sim 560$ GeV at $M'_1\sim 800$ GeV and then
falls down crossing all possible $s$-channel resonances
twice. Starting from $M'_1=0$ we first arrive at a little dip in the
relic density around $M'_1=250$ GeV due to the $s$-channel
$H_2/A$ resonance, and a little wiggle around
$M'_1=500$ GeV due to $Z_2/H_3$  as the LSP has not yet
developed an appreciable singlino component.  The first appreciable
dip in the relic density occurs around $M_1'=0.8$~TeV where
$\Omega_{CDM}h^2$ drops to $\sim 0.02$. Here the LSP has a strong
higgsino component which enhances the annihilation via the s-channel
$Z_2/H_3$ resonances considerably. Increasing $M'_1$ further, the LSP
mass increases, going off-resonance (hence local maximum in the relic
density), until it reaches its maximum of $~\sim 590$ GeV at $M'_1\sim
800$ GeV. From now on the LSP mass decreases and its nature becomes
singlino-dominated. Around $M'_1=1.5$ TeV it once again hits the
$Z_2/H_3$ resonance.  However, this time the LSP is predominantly
singlino. Although pure singlino neutralinos do not couple to the
singlet Higgs, so the $H_3$ resonance is subdominant, they couple
strongly to the $Z'$ and annihilate very efficiently.  As a result,
the relic density drops to $\sim 2\times 10^{-3}$.
The kink at $M_1'=2.5$~TeV develops as the LSP mass drops below threshold for production of $H_1A$
in the final state. Increasing $M'_1$ further we find a pseudoscalar Higgs resonance at
$M_1'=3.5$~TeV, the top threshold at $M_1'=5$~TeV, the light Higgs
threshold at $M_1'=11$~TeV, the light Higgs resonance at $M_1'=14$~TeV
and the $Z$ resonance at $M_1'=19$~TeV.

\subsection{Direct detection}
\label{scenB-dd}
Compared to the MSSM, the spin-dependent elastic scattering receives an additional contribution due to a heavy $Z_2$ gauge boson, while the spin-independent one gets contributions from additional Higgs as well as from interactions generated by the $g'_1\tilde B'(\tilde H_i H_i+
\tilde S S)$  and  $\lambda\tilde H_i (\tilde S H_j + \tilde
H_j S)$ couplings.
\begin{figure}[h]
\centerline{\includegraphics[width=0.45\columnwidth, height=0.3\columnwidth]{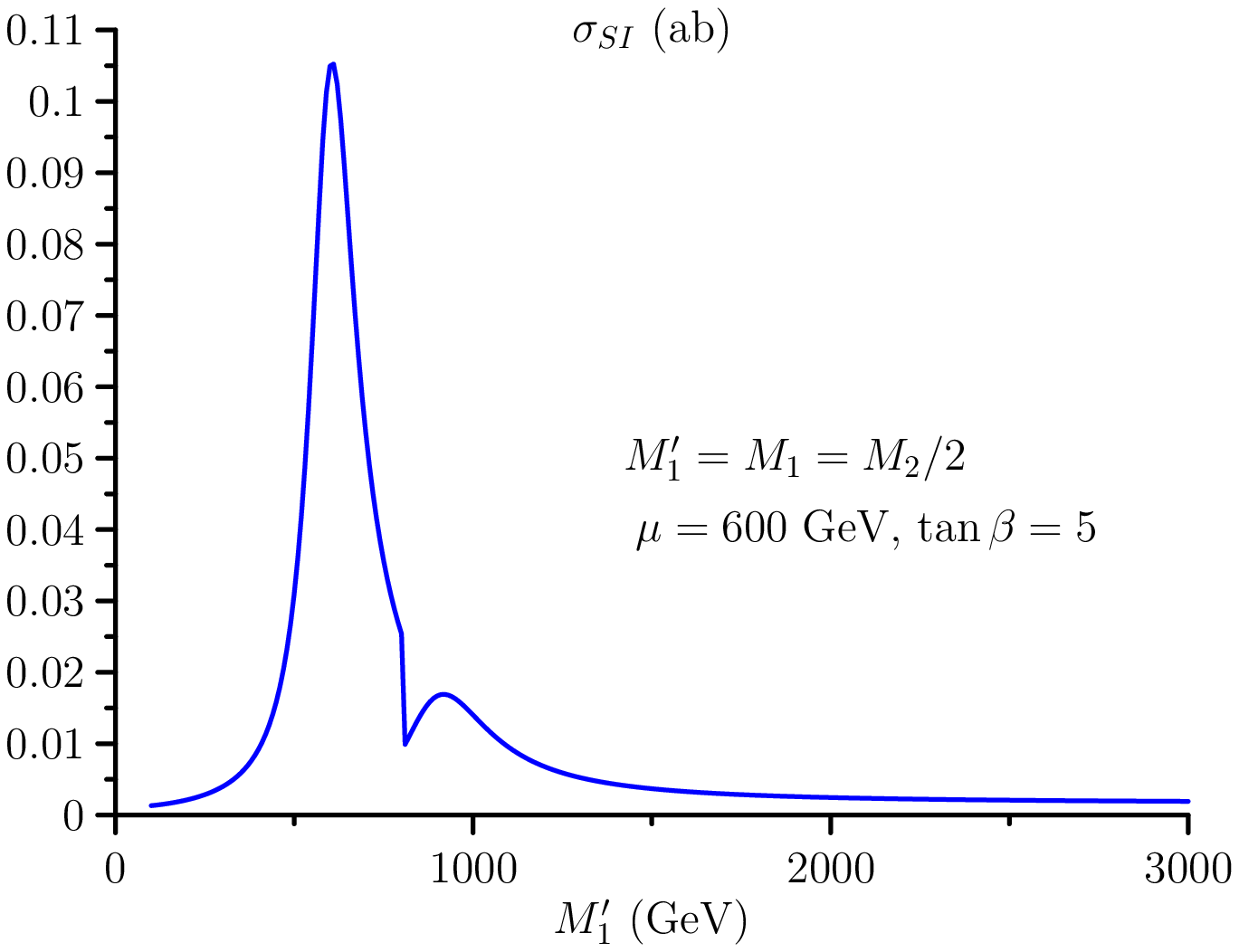}~~
\includegraphics[width=0.45\columnwidth, height=0.3\columnwidth]{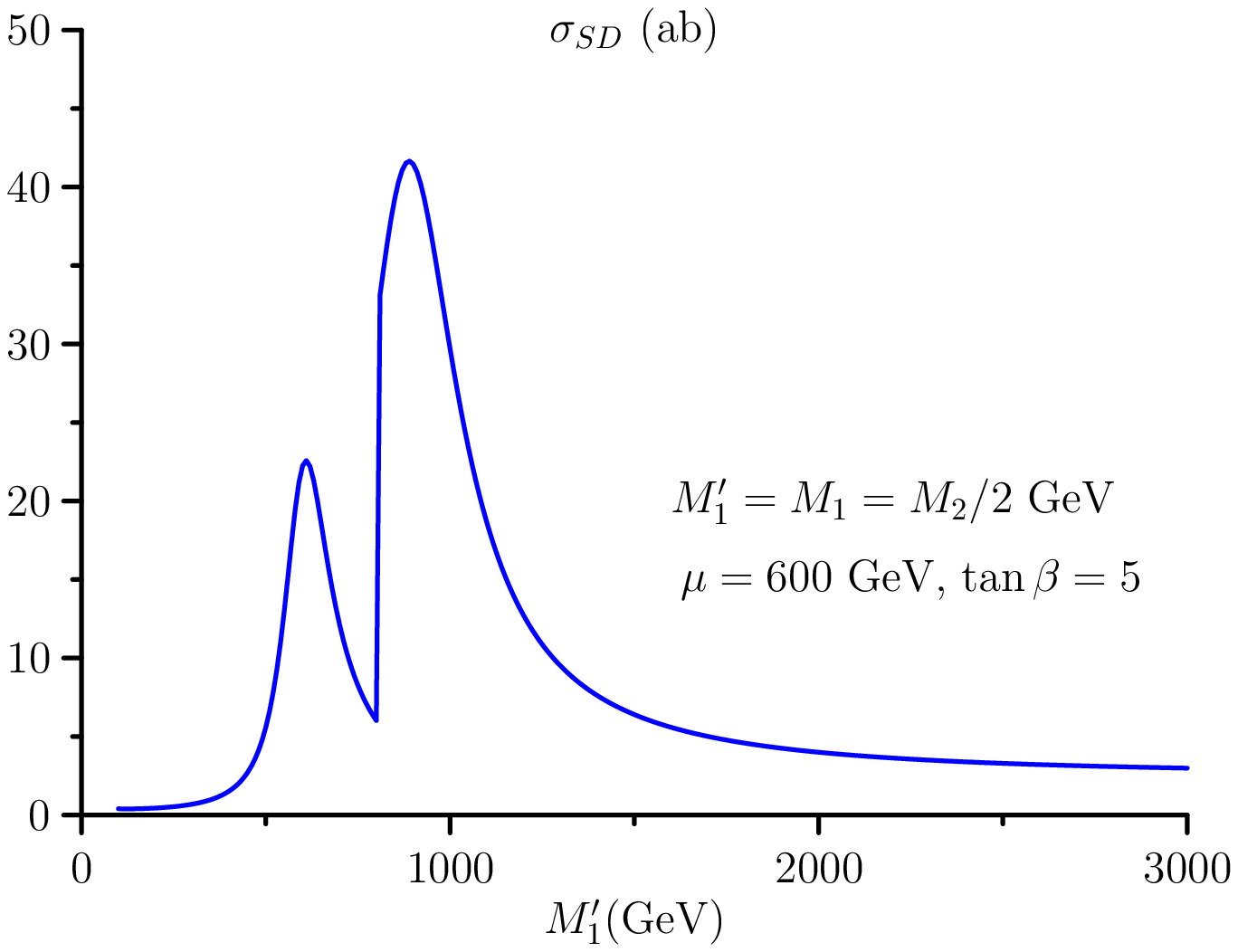}}
\caption{The elastic spin-independent (left) and spin-dependent
  (right) LSP-proton cross section as a function of $M_1'$ }\label{Fig:direct}
\end{figure}
In Figure~\ref{Fig:direct} the spin-independent as well as
spin-dependent elastic cross section of the lightest neutralino on the
proton is shown as a function of $M'_1$
(restricted to 0--3~TeV, as beyond this range cross sections fall
monotonically). Referring to Fig.~\ref{Fig:masses}, it is easy to understand
the $M_1'$ behavior of the cross section. For $M'_1$ below 800 GeV, the
LSP is mostly a mixture of the MSSM states. It starts as a
bino  and as $M'_1$
approaches 500 GeV, it receives an appreciable admixture of both
higgsinos. As a result both $\sigma_{SD}$ and $\sigma_{SI}$ first rise and then fall. The fall is due to the diminishing bino component, which reduces the Higgs contribution to $\sigma_{SI}$, and increases the  cancelation in $c_{34}=|N_{13}|^2-|N_{14}|^2$, that controls the $Z$ contribution to
$\sigma_{SD}$. At
$M'_1\sim 800$~GeV the discontinuities reflect crossing of two lightest states
since the mixing with singlino forces the second-lightest state to become the lightest.
At this point $\sigma_{SI}$ drops by a factor 3. As the singlino
component of the LSP increases with $M'_1$ the $\lambda\tilde H_i \tilde S H_j$ term
becomes responsible for the little rise of the
cross section. On the other hand, $\sigma_{SD}$ jumps by a factor 6 at the discontinuity since
the mixing with singlino upsets the cancelation in $c_{34}$. With further increase of $M'_1$ the LSP becomes almost a pure singlino which explains a steady fall of both cross sections.

\section{Summary}
The USSM, despite its modest additional particle
content compared to the MSSM or NMSSM, leads to a surprisingly rich
and interesting dark matter phenomenology which distinguishes it from
these models. There are many cases where successful relic abundances may be reached,
either through a proper balance of the singlino/higgsino mixture, or
through a balance of the singlino mass against the mass of a
boson that mediates annihilation in the s-channel. The difference in the Higgs
spectrum and the singlino interactions results in significant
differences in the direct detection predictions as well.

\section*{Acknowledgments}
Work partially supported by the Polish MNiSW Grant No~1~P03B~108~30, the EC Programmes MTKD--CT--2005--029466 and MRTN-CT-2006-035505, as well as   STFC
Rolling Grant PPA/G/S/2003/00096, EU Network MRTN-CT-2004-503369, EU
ILIAS RII3-CT-2004-506222, NSF CAREER grant PHY-0449818 and DOE OJI
grant \# DE-FG02-06ER41417.


\begin{footnotesize}


\end{footnotesize}


\begin{thebibliography}{99}


\bibitem{Fayet:1977yc}
  P.~Fayet,
  Phys.\ Lett.\  B {\bf 69} (1977) 489.


\bibitem{King:2005jy}
  S.~F.~King, S.~Moretti and R.~Nevzorov,
  Phys.\ Rev.\  D {\bf 73} (2006) 035009.


\bibitem{our}
D.~Jarecka, J.~Kalinowski, S.~F.~King and J.~P.~Roberts,
{\it Proc.\ of LCWS07 and ILC07, Hamburg, Germany, 2007, pp SUS15}
  [arXiv:0709.1862 [hep-ph]];
  J.\ Phys.\ Conf.\ Ser.\  {\bf 110}, 072019 (2008).


\bibitem{Kalinowski:2008iq}
  J.~Kalinowski, S.~F.~King and J.~P.~Roberts,
  arXiv:0811.2204 [hep-ph].

\bibitem{Choi:2006fz}
  S.Y.~Choi, H.E.~Haber, J.~Kalinowski and P.M.~Zerwas, Nucl.\ Phys.\ B
  {\bf 778} (2007) 85.



\bibitem{micromegas}
  G.~Belanger, F.~Boudjema, A.~Pukhov and A.~Semenov,
  Comput.\ Phys.\ Commun.\  {\bf 176}, 367 (2007).



\end{thebibliography}
\end{document}